\documentclass[12pt]{article}
\usepackage{amsmath}
\usepackage{amssymb}
\usepackage{amsfonts}
\usepackage{graphicx}

\newcommand{\ov } {\over }
\newcommand{\be}{\begin{equation}}
\newcommand{\ee}{\end{equation}}
\newcommand{\bea}{\begin{eqnarray}}
\newcommand{\eea}{\end{eqnarray}}

\begin{document}

\bigskip 
\begin{titlepage}
\begin{flushright}
UUITP-20/08\\

\end{flushright}
\vspace{1cm}
\begin{center}
{\Large\bf Chain inflation and the imprint of fundamental physics in the CMBR\\}
\end{center}
\vspace{3mm}
\begin{center}
{\large
Diego Chialva$^1$ and Ulf H.\ Danielsson$^2$} \\
\vspace{5mm}
Institutionen f\"or fysik och astronomi \\
Uppsala Universitet,
Box 803, SE-751 08
Uppsala, Sweden

\vspace{5mm}
 
{\tt
{$^1$}diego.chialva@fysast.uu.se \qquad
{$^2$}ulf.danielsson@fysast.uu.se \\
}
\end{center}
\vspace{5mm}

\begin{center}
{\large \bf Abstract}
\end{center}
\noindent
In this work we investigate characteristic modifications of the
spectrum of cosmological perturbations and the spectral index due to chain inflation.
We find two types of effects. First, modifications of the spectral index depending on interactions
between radiation and the vacuum, and on features of the effective
vacuum potential of the underlying fundamental theory. Second,
a modulation of the spectrum signaling new physics due to bubble nucleation. This
effect is similar to those of
transplanckian physics.
Measurements of such signatures could provide a wealth of information on the
fundamental physics at the basis of inflation.
\vfill
\begin{flushleft}
September 2008
\end{flushleft}
\end{titlepage}\newpage

\tableofcontents


\section{Introduction}


\bigskip

The current most popular models for inflation are based on chaotic inflation.
In these models a scalar field rolls slowly subject to Hubble friction in a
shallow potential. In \cite{Chialva:2008zw} we proposed an alternative
scenario\footnote{For earlier attempts on models featuring inflation through
chain of first order decays, see also \cite{Freese:2004vs, Freese:2006fk,
Huang:2007ek, Huang:2008jr}} that shares many of the features of slow roll
chaotic inflation, but also differs in several important aspects. Our model is
based on chain inflation and makes use of a series of several first order
phase transitions. More precisely, we imagine a potential with a large number
of small barriers that separate local, metastable minima. The barriers prevent
the field from rolling, and without quantum mechanical tunneling the inflaton
is stuck in a local minimum. By appropriately choosing the heights and widths
of the barriers, one can obtain tunneling probabilities such that the field
effectively rolls slowly down the potential through repeated tunneling events.
In this way we can achieve a slow roll in the sense of having a slow change in
$H^{2}\sim\rho^{V}$ ($\rho^{V}$ being the vacuum energy density), even if the
potential for the fields is steep. The details of this process were worked out
in \cite{Chialva:2008zw}, and it was also shown that suitable potentials might
be find in flux compactified string theory.

The main features of the model introduced in \cite{Chialva:2008zw} is as
follows. We assume that the bubbles, after their formation through tunneling,
rapidly percolate and collide. The energy difference between two subsequent
minima is temporarily stored in the bubble walls, and we assume that this
energy is rapidly converted into radiation as the bubbles collide. In this way
we obtain a coarse grained picture where the main effect of the barriers and
the tunneling is to introduce a source term for radiation in the Friedmann equations.

A scalar field can be understood as a fluid consisting of two components: a
component corresponding to the kinetic energy, $T$, and a component
corresponding to the potential energy, $V$. In the case of slow roll we have
$T\sim\varepsilon V\ll V,$ where $\varepsilon$ is the slow roll parameter, and
as a consequence the dynamics is dominated by the potential energy leading to
accelerated expansion and inflation. In our version of chain inflation the
kinetic component is further suppressed relative to the potential energy. On
the other hand, radiation is produced through the tunneling leading to
$\rho_{rad}\sim\varepsilon V$. As a result we effectively have, to first order
in $\varepsilon$, a model consisting of a decaying cosmological constant and a
coupled component of radiation. For the case of chaotic inflation, it is
important to understand that it is the sub-dominant kinetic energy that
determines the spectrum of the fluctuations. The kinetic energy corresponds to
a hydrodynamical fluid with an effective speed of sound that is equal to the
speed of light. In contrast, the potential energy does not correspond to a
hydrodynamical fluid and lacks a well defined speed of sound.

The amplitude of the primordial fluctuations are set by the speed of sound.
The general result is
\begin{equation}
P\sim\frac{H^{2}}{c_{s}\varepsilon},
\end{equation}
where $c_{s}$is the speed of sound of the hydrodynamical component. For
chaotic inflation $c_{s}=1$. In our model for chain inflation, the role of the
kinetic energy is taken over by the radiation where $c_{s}=\frac{1}{\sqrt{3}}%
$. As a result,the primordial spectrum is corrected to
\begin{equation}
P\sim\sqrt{3}\frac{H^{2}}{\varepsilon}.
\end{equation}
The result differs from chaotic inflation through a simple factor of $\sqrt
{3}$. While this simple argument captures the main physics of the model, there
are many important points of the derivation that are carefully discussed in
\cite{Chialva:2008zw}.

In the present paper we discuss the possibility of further effects on the
primordial spectrum from various sources.\footnote{We will leave aside the
interesting possibility of having sizable contributions from isocurvature
perturbations and/or non-gaussianities.} In the first part of the paper we
will discuss the modifications to the spectrum of cosmological perturbations
due to the presence of the non-negligible interaction between radiation and
vacuum energy. We will discuss how they arise and appear to be specific to our
model of chain inflation. In the second part of the paper we will instead
consider how bubble nucleation affect the spectrum of perturbations, and in
particular we will study the imprint of the size of the bubbles on the CMB.


\section{Effects due to interactions}


In \cite{Chialva:2008zw} we derived a system of equations that determine the
evolution of the comoving curvature perturbations during a period of chain
inflation. The approach was based on the traditional analysis of scalar
perturbations in field (slow-roll/chaotic) models, as presented in
\cite{kinflGarrMuk}. We start with a brief review of the approach used in
\cite{Chialva:2008zw}, and show that the method of \cite{kinflGarrMuk} is not
the most convenient one in the case of chain inflation. We will therefore
propose another way of analyzing and re-writing the system of equations that
is better suited for our model.

We start from equation (97) in \cite{Chialva:2008zw},
\begin{equation}
\left\{
\begin{aligned} \dot\xi & = {a(\rho+p) \ov H^2}\mathcal{R} \\ \dot{\mathcal{R}} & = {1 \ov 3} {H^2 \ov a^3 (\rho+p)}\big(-k^2 -a^24\pi G {Q_V \ov 3H}\big)\xi -{4 \ov 3} H S_{V, r} \end{aligned}\right.
, \label{systpert}%
\end{equation}
where

\begin{itemize}
\item $a$ is the cosmological scale factor, $H$ is the Hubble rate, $\phi$ is
the gravitational potential, and $G$ is Newton's constant

\item $Q_{V}$ is the energy-momentum transfer vector\footnote{See
\cite{Chialva:2008zw} for its exhaustive definition. For what we are concerned
it is sufficient to define it through the equations
\begin{align}
\dot{\rho}^{V}  &  =Q_{V}\\
\dot{\rho}^{r}  &  =-4H\rho^{r}-Q_{V}%
\end{align}
where $\rho^{V/r}$ is the energy density for vacuum/radiation.}

\item $a\phi= 4\pi GH\xi$,

\item $\mathcal{R}$ is the comoving curvature perturbation

\item $\rho,\,p$ are the total energy and pressure density

\item $\varepsilon$ is the slow-roll parameter

\item $k$ is the comoving wavenumber for the perturbation

\item $S_{V,r}$ is the relative entropy perturbation between vacuum ($V$) and
radiation ($r$).
\end{itemize}

In the following we will neglect the term proportional to $S_{V,r}$. As a
result our conclusions apply only to models with negligible contributions from
isoentropic perturbations, or, alternatively, just to the adiabatic component
of the whole spectrum of perturbations.

Comparing equations (\ref{systpert}) with the analogous equations in
\cite{kinflGarrMuk} (in flat space), we see the importance of interactions in
our multicomponent system, as represented by the term $-a^{2}4\pi
G{Q_{V}\ov3H}\xi$. Note also that this can be conveniently re-written as
$-a^{2}4\pi G{Q_{V}\ov3H}=a^{2}H^{2}\varepsilon$. Following the literature
(see for example \cite{Mukhanov:1990me}), it is customary to define a standard
quantization variable $\varsigma=z\mathcal{R}$ where
\begin{equation}
z\equiv{a(\rho+p)^{1\ov2}\ov {1 \ov 3}H}\left(  {\hat{O}\ov(-k^{2}+a^{2}%
H^{2}\varepsilon)}\right)  ^{1\ov2}.
\end{equation}
The singularity at $k^{2}=a^{2}H^{2}\varepsilon$ is just an artifact of the
choice of variables, as is evident from (\ref{systpert})\footnote{This issue
is absent in \cite{kinflGarrMuk} in the case of flat space}. However, in order
to better understand the implications for the spectrum of perturbations due to
the new term, we choose to follow an alternative route using another change of
variables. The (first order) action inferred from the equations of motion is
given by\footnote{This is the same action as in \cite{kinflGarrMuk} with the
derivative term integrated by parts.}
\begin{equation}
S=\int\big[\dot{\xi}\hat{O}\mathcal{R}+{1\ov2}{H^{2}c_{s}^{2}\ov a^{3}%
(\rho+p)}\xi(\Delta+a^{2}H^{2}\varepsilon)\hat{O}\xi-{1\ov2}{a^{2}%
(\rho+p)\ov H^{2}}\mathcal{R}\hat{O}\mathcal{R}\big]dt\,d^{3}x,
\end{equation}
where $\hat{O}$ is a time-independent factor which, by comparison with known
cases, was found to be\footnote{In our previous paper we used $\hat{O}=-k^{2}%
$, but here we prefer use the other sign.} $\hat{O}=k^{2}$. By eliminating
$\mathcal{R}$ in the action by using the first equation in (\ref{systpert}),
and defining
\begin{align}
w^{2}  &  \equiv{H^{2}\ov(\rho+p)a^{2}}\hat{O}\\
u  &  \equiv w\,\xi\,,
\end{align}
we obtain the action (up to a total derivative term)
\begin{equation}
S={1\ov2}\int\big[u^{^{\prime}2}+{1\ov3}u(\Delta+\mathcal{H}^{2}%
\varepsilon)u+{w^{^{\prime\prime}}\ov w}u^{2}\big]d\!\eta\,d^{3}\!x,
\end{equation}
and the equation of motion is
\begin{equation}
u^{^{\prime\prime}}+\left(  {1\ov3}k^{2}-{1\ov3}\mathcal{H}^{2}\varepsilon
-{w^{^{\prime\prime}}\ov w}\right)  u=0.
\end{equation}
Here we have used the conformal time defined as $\eta=\int a^{-1}dt$ and
$\mathcal{H}={a^{\prime}\ov a}$, where a prime represents a derivative with
respect to the conformal time. An advantage with this way of writing the
equations is the absence of any artificial singularities.

\subsection{The equation of motion}

We will now expand ${w^{^{\prime\prime}}\ov w}$ in slow-roll parameters such
as $\varepsilon$ and the decay rates in such a way that we make sure to
include contributions from derivatives of the slow-roll parameters up to order
$O(\varepsilon)$. We will study the general case of a vacuum energy
$\epsilon_{n}$ in the form of a power law,
\begin{equation}
\epsilon_{n}\sim{m_{f}^{4}\ov c!}n^{c}, \label{powerlaw}%
\end{equation}
where the integer $n$ labels the vacuum, and $m_{f}$ is an energy scale
(depending on couplings, extra-dimensions and similar features of the
microscopical theoretical model). We will need the following formula (see
appendix for derivation)
\begin{equation}
\dot{\varepsilon}\sim\left(  1+{2\ov c}\right)  H\varepsilon^{2}%
-D\widetilde{\Gamma}, \label{derivslowrollfinal}%
\end{equation}
where $D\widetilde{\Gamma}$ depends on the decay rates per unit time
$\widetilde{\Gamma}$ and is defined in the appendix. Note that we do not
restrict to equal rates at every step. We find
\begin{equation}
{w^{\prime\prime}\ov w}\sim\mathcal{H}^{2}\varepsilon+{1\ov2}\mathcal{H}%
^{2}\left(  \big(1+{2\ov c}\big)\varepsilon-{\ D\widetilde{\Gamma}%
\ov H}\right)
\end{equation}
having neglected terms of order $\varepsilon^{2},\varepsilon D\widetilde
{\Gamma},D\widetilde{\Gamma}^{2}$ and higher. Our equation of motion now becomes%

\begin{equation}
u^{^{\prime\prime}}+\left(  {1\ov3}k^{2}-{1\ov3}\mathcal{H}^{2}\varepsilon
-\mathcal{H}^{2}\varepsilon-{1\ov2}\left(  \left(  1+{2\ov c}\right)
\mathcal{H}^{2}\varepsilon-a\mathcal{H}D\widetilde{\Gamma}\right)  \right)
u=0.
\end{equation}
In a quasi-deSitter space, where
\begin{equation}
a\sim-{1\ov H\eta(1-\varepsilon)}\qquad(\eta<0),
\end{equation}
we have
\begin{equation}
{1\ov\eta^{2}}\left(  \nu^{2}-{1\ov4}\right)  =\mathcal{H}^{2}\left(
{\varepsilon\ov3}+\varepsilon+{1\ov2}\left(  \left(  1+{2\ov c}\right)
\varepsilon-{D\widetilde{\Gamma}\ov H}\right)  \right)  ,
\end{equation}
which allow us to read off
\begin{equation}
\nu^{2}\sim{1\ov4}+{\varepsilon\ov3}+\varepsilon+{1\ov2}\left(  \left(
1+{2\ov c}\right)  \varepsilon-{D\widetilde{\Gamma}\ov H}\right)  .
\end{equation}

The general solution for the equation
\begin{equation}
u^{^{\prime\prime}}+\left(  {1\ov3}k^{2}-{1\ov\eta^{2}}\left(  \nu^{2}%
-{1\ov4}\right)  \right)  u=0
\end{equation}
reads
\begin{equation}
u=\sqrt{-\eta}\left[  c_{1}(k)H_{\nu}^{(1)}(-k\eta)+c_{2}(k)H_{\nu}%
^{(2)}(-k\eta)\right]  ,
\end{equation}
where $H_{\nu}^{(1)}(x)$ and $H_{\nu}^{(2)}(x)$ are Hankel functions of the
first and second kind, respectively. In the limit $-k\eta\gg1$ we have that
\begin{equation}
H_{\nu}^{(1)}\sim\sqrt{{2\ov-\pi k\eta}}e^{i(-k\eta-{\pi\ov2}\nu-{\pi\ov4}%
)}\qquad H_{\nu}^{(2)}\sim\sqrt{{2\ov-\pi k\eta}}e^{-i(-k\eta-{\pi\ov2}%
\nu-{\pi\ov4})}\,\,.
\end{equation}
Following standard procedure we match the solution with the Bunch-Davies
vacuum\footnote{In the following section we will discuss this choice thorough,
investigating the possibility of new fundamental physics at a scale larger
than the Plank one.} ${e^{-ik\eta}\ov\sqrt{2kc_{s}}}$, finding
\begin{equation}
u={1\ov2}\sqrt{{\pi\ov c_{s}}}e^{i(\nu+{1\ov2}){\pi\ov2}}\sqrt{-\eta}H_{\nu
}^{(1)}(-k\eta).
\end{equation}
For superhorizon scales ($-k\eta\ll1$)
\begin{equation}
H_{\nu}^{(1)}\sim\sqrt{{2\ov\pi}}e^{-i{\pi\ov2}}2^{\nu-{3\ov2}}{\Gamma
(\nu)\ov\Gamma({3\ov2})}(-k\eta)^{-\nu}%
\end{equation}
so that, finally,
\begin{equation}
u\sim e^{i{\pi\ov2}(\nu-{1\ov1})}2^{\nu-{3\ov2}}{\Gamma(\nu)\ov\Gamma
({3\ov2})}\frac{{(-k\eta)^{{1\ov2}-\nu}}}{\sqrt{2kc_{s}}}{.}%
\end{equation}
This is what we need in order to obtain the spectrum of perturbations.

\subsection{Spectrum of perturbations}

The spectrum of perturbations is conveniently expressed through the comoving
curvature perturbation, which is constant on superhorizon scales during
inflation. We can obtain it using the first equation in (\ref{systpert}),
which we repeat here for convenience
\begin{equation}
\mathcal{R}={H^{2}\ov a(\rho+p)}\dot{\xi}.
\end{equation}
As a result we obtain
\begin{equation}
\mathcal{R}={H\ov\sqrt{2M_{\text{Plank}}\varepsilon}}{1\ov\sqrt{2k^{3}c_{s}}%
}\left(  {k\ov aH}\right)  ^{{1\ov2}-\nu}(1+O(\varepsilon)),
\end{equation}
and the spectrum becomes
\begin{equation}
P_{k}^{\mathcal{R}}={H^{2}\ov8\pi^{2}M_{Plank}^{2}{1\ov\sqrt{3}}\varepsilon
}\left(  {k\ov aH}\right)  ^{1-2\nu}(1+O(\varepsilon))\,.
\end{equation}
From this we read off the spectral index
\begin{equation}
\label{spectindexeps}n_{s}-1=1-2\nu\sim-{\ 2\ov3}\varepsilon-2\varepsilon-
\left(  1+{2\ov c}\right)  \varepsilon+{D\widetilde{\Gamma}\ov H,}%
\end{equation}
which alternatively can be written as
\begin{equation}
\label{spectindexQV}n_{s}-1\sim{c_{s}^{2}\ov3H^{3}M_{Plank}^{2}}%
Q_{V}-2\varepsilon- \left(  1+{2\ov c}\right)  \varepsilon+\frac
{{D\widetilde{\Gamma}}}{{H}}{,}%
\end{equation}
which is our final result. It is evident from this formula that the
corrections to the spectral index due to the interactions between vacuum and
radiation imply an extra tilt to the spectrum (blue or red depending
on the value of $D\widetilde{\Gamma}$). (Note that the result for
$D_{\sigma}\widetilde{\Gamma}=Q_{V}=0$ is precisely the same as the usual
chaotic/slow-roll result in the case of $c=2$ as observed in
\cite{Chialva:2008zw}.) It appears to us that this feature of the spectrum is
strongly characteristic of a first order transition, since in deriving
(\ref{systpert}) in \cite{Chialva:2008zw}, we made use of specific aspects of
first order transitions (such as the fact that the momentum perturbation for
the vacuum was zero).


\section{Effects of a new intermediate scale}


\bigskip

\subsection{The choice of vacuum}

In deriving our results for the spectrum of cosmological perturbations and the
spectral index in the previous section, we have followed the standard
procedure of matching our solution to the Bunch-Davies vacuum. Essentially
this means that we resolve the issue of the non-uniqueness of the vacuum in a
cosmological space time by tracking the modes to infinitely short scales,
where the effect of cosmological scales such as the horizon can be ignored. At
such scales there is a unique vacuum just as in Minkowsky space. This is the
Bunch-Davies vacuum.

As is well known there is a potential problem with this procedure since one
can not reliably track the modes to scales shorter than the Planck (or string)
scale without taking into account effects of string theory and quantum
gravity\footnote{See for example \cite{Martin:2000xs, Niemeyer:2000eh,
Danielsson:2002kx, Danielsson:2002qh, Easther:2002xe}}. Hence there are likely
corrections to the choice of vacuum of order $\frac{H}{\Lambda}$, where
$\Lambda$ is the scale of new physics. This is known as the transplanckian
problem. Actually, it represents more of an opportunity than a problem since
it could be an observational window to new physics.

In our case we have yet another scale that enters. We have assumed a coarse
graining over the nucleating bubbles that is valid only for scales
substantially larger than the size of the bubbles, $r_{b}$. Hence, we have
full control over the evolution of the perturbations only while their
wavelength is larger than the size of the bubbles. If we follow the evolution
of a specific mode backwards in time it will eventually reach a scale as short
as the size of the bubbles, and our picture breaks down.

What is the effective quantum state that should be used as an initial
condition at this point? It is in general very difficult to give a precise
answer to this question both because of the usual difficulties due to
quantization in curved spaces, and also because of the great generality of our
model of chain inflation (no field theory model is specified). Without a
detailed model, we have no other option than to impose an effective initial
condition for the perturbations, and to postulate their creation out of the
vacuum. This is formally very similar to the case of the transplanckian
problem. In several works (\cite{Martin:2000xs, Niemeyer:2000eh,
Danielsson:2002kx, Danielsson:2002qh, Easther:2002xe}) it has been argued that
initial conditions must be imposed at the Planck scale (or string scale) due
to our ignorance of physics at higher energies. In our model the scale for new
physics will instead be the size of the bubbles, but the analysis, that we now
review, will be more or less the same.

We begin by noting that the physical momentum $p$ and the comoving momentum
$k$ are related through
\begin{equation}
k=ap=-\frac{p}{\eta H},
\end{equation}
where $\eta$ is conformal time, $p$ is the physical momentum, $k$ is the
comoving momentum and $a$ is the scale factor. We impose the initial
conditions when $p=\Lambda$, where $\Lambda$ is the energy scale important for
the new physics given by $\Lambda\sim1/r_{b}$. In our case $r_{b}$ is just the
size of the bubbles. We find that the conformal time when the initial
condition is imposed to be
\begin{equation}
\eta_{0}=-\frac{\Lambda}{Hk}.
\end{equation}
As we see, different modes will be created at different times, with a smaller
linear size of the mode (larger $k$) implying a later time.

In our case we would in principle be able to calculate the form of the
perturbations at the the scale $r_{b},$ by tracing the evolution backwards in
time, through the nucleating bubbles, to even smaller scales. Presumably the
result would depend on the fine details of the physics of bubble nucleation,
which is beyond the scope of the present paper.

Instead we will take the same attitude as in \cite{Danielsson:2002kx} and
encode the unknown new physics into the choice of the vacuum. The claim of
\cite{Danielsson:2002kx} is that the primordial spectrum is corrected through
a modulating factor. These results can be directly taken over to our case with
the result that
\begin{equation}
\label{spectrmod}P\left(  k\right)  \sim\frac{H^{2}}{\varepsilon c_{s}}\left(
1-\frac{H}{\Lambda}\sin\left(  \frac{2\Lambda}{H}\right)  \right)  ,
\end{equation}
where $c_{s}=\frac{1}{\sqrt{3}}$ is the speed of sound. In the transplanckian
case, $\Lambda$ is typically constant and equal to the Planck scale or string
scale. The modulation of the spectrum comes from the rolling inflaton that
leads to a changing $H$. In our case, $\Lambda$ will also be changing, but the
amplitude $\frac{H}{\Lambda}$ is nevertheless expected to be small, and not to
change very much during inflation. The argument of the sine, on the other
hand, is a large number and can easily change by several times $2\pi$ during
the relevant time period for the generation of the primordial perturbations.

Let us now investigate in more detail what the effect will be in the case of
chain inflation.

\subsection{The size of the bubbles}

To proceed we need to know more about the process of nucleation of bubbles.
The rate of bubble formation per unit time and physical volume, in the
analysis of \cite{Coleman:1977py}, \cite{Coleman:1980aw}, is given by the
exponential of the euclidean instanton action, $S_{E},$ responsible for the
tunneling (the so-called \textquotedblleft bounce"),
\begin{equation}
\Gamma\sim e^{-S_{E}}.
\end{equation}
The action evaluated on the bounce is given by\footnote{We recognize in this
the variation of the Gibbs energy for the nucleation of a bubble: with a
different normalization for what concerns energy density and surface
tension.}:
\begin{equation}
S_{E}=-{\pi^{2}\ov2}r^{4}\Delta\epsilon+2\pi^{2}r^{2}S,
\end{equation}
where $r$ is the radius of the bubble, $\Delta\epsilon$ the change in energy
due to the nucleation of the bubble, and $S$ is the bubble's wall
tension.\footnote{If the tension is due to a scalar field, $\phi,$ we have
that $S=\int d\phi\sqrt{V}$, where $V$ is the potential.} In principle, there
is also a third term present due to the effect of gravity, but we assume the
size of the bubbles to be much smaller than the Hubble scale so that we can
ignore it. The critical radius that allows for the nucleation of a bubble that
will successfully expand, and therefore enables tunneling, is obtained by
extremizing the above Euclidean action. The result is%
\begin{align}
r_{b}\equiv r_{\text{critical}}  &  =\frac{3S}{\Delta\epsilon},\\
S_{E}\big|_{r=r_{\text{critical}}}  &  =\frac{27\pi^{2}}{2}\frac{S^{4}%
}{\left(  \Delta\epsilon\right)  ^{3}}.
\end{align}

The setup outlined in these formulas is a static one. The tunneling occurs
between two vacua of the theory, and any possible time evolution of the
background is not taken into account. In our scenario, on the other hand, we
have a chain of tunneling events occurring through time. The length scale
signaling new physics (corresponding to the radius of the nucleated bubbles)
depends on the time when the particular mode of interest is produced. For
simplicity, however, we will assume that the change in the radius is slow
enough that we can use the above analysis.

Accounting for the time evolution of the background, when computing the
critical radius of the bubbles at a given time, is most easily achieved by
expressing the variation of the energy density due to the nucleation as
\begin{equation}
\Delta\epsilon\sim-{d\rho^{V}\ov dt}\langle\tau\rangle=6H^{3}M_{\text{Plank}%
}^{2}\langle\tau\rangle\,\varepsilon,
\end{equation}
where we have used the Friedman equation, and defined $\langle\tau
\rangle\equiv\langle{\widetilde{\Gamma}}\rangle^{-1}$ to be the average
tunneling time (we recall that $\widetilde{\Gamma}$ is the decay rate per unit
time)\footnote{All averages are taken with the distribution of vacuum phases
$\rho_{m}^{V}=\epsilon_{m}p_{m}(t)$, see appendix and \cite{Chialva:2008zw}}.
Also, the surface tension $S$ needs to have a time dependence. It is more
convenient, though, to express $S$ through the extremized action, as
\begin{equation}
S=\left(  {2S_{E}\ov27\pi^{2}}\right)  ^{{1\ov4}}\Delta\epsilon^{{3\ov4}},
\end{equation}
Eventually we find
\begin{align}
r_{b}H  &  =\left(  2S_{E}\ov c\pi^{2}\right)  ^{{1\ov4}}\langle\tau
\rangle^{-{1\ov4}}\langle{\widetilde{\Gamma}\ov n}\rangle^{-{1\ov4}}\left(
\frac{H}{M_{pl}}\right)  ^{{1\ov4}}\nonumber\\
&  =\left(  2S_{E}\ov c\pi^{2}\right)  ^{{1\ov4}}\langle\tau\rangle^{-{1\ov4}%
}\langle{\widetilde{\Gamma}\ov n}\rangle^{-{1\ov4}}\left(  8\pi^{2}%
\,\eta\,{\varepsilon\ov\sqrt{3}}\right)  ^{{1\ov4}},
\end{align}
where $\eta\sim2.5\cdot10^{-9}$ from the normalization of the spectrum. With
$\varepsilon=10^{-2}$ we find
\begin{equation}
r_{b}H\sim3.9\cdot10^{-3}S_{E}^{{1\ov4}}c^{-{1\ov4}}\langle\tau\rangle
^{-{1\ov4}}\langle{\widetilde{\Gamma}\ov n}\rangle^{-{1\ov4}}.
\end{equation}
In our calculation we have ignored time dependent corrections to, e.g.,
${S_{E}}$ that are suppressed by $1/n$.

Let us now consider the possible observational implications of the above
effects. Successful chain inflation requires that while the vacuum undergoes
the transitions, the phase distribution in the universe is peaked
consecutively on the various phases. That is, the transitions occur
consecutively, and in a short time (shorter than the Hubble time). Rapid
tunneling implies that ${S_{E}}$ should be at most of order one, in order for
$\langle\tau\rangle H\ll1$. If we then use $\varepsilon=\frac{c}{2H}%
\langle{\widetilde{\Gamma}\ov n}\rangle=\frac{c}{2\langle\tau\rangle H}%
\langle\tau\rangle\langle{\widetilde{\Gamma}\ov n}\rangle,$ we find that
$\langle\tau\rangle\langle{\widetilde{\Gamma}\ov n}\rangle$ needs to be small.
With at peaked distribution we have, to a good approximation, $\langle
\tau\rangle\langle{\widetilde{\Gamma}\ov n}\rangle\sim1/n,$ and we see that
$n$ needs to be large.

Turning back to equation (\ref{spectrmod}), and the corrections to the
spectrum from the presence of a new scale, we know from the work on
transplanckian physics that values of the order $\varepsilon\sim10^{-2}$ and
$\frac{H}{\Lambda}\sim10^{-3}$ could possibly yield an observational effect.
The restriction on $\varepsilon$ comes from the requirement that $H$ changes
in an appreciable way in order for there to be a modulation. Using $H\sim
k^{-\varepsilon}$ and (\ref{spectrmod}) we have, following
\cite{Bergstrom:2002yd},
\begin{equation}
\frac{\Delta k}{k}\sim\frac{\pi H}{\varepsilon\Lambda},
\end{equation}
where, in our case, $\Lambda\sim1/r_{b}$. We see that $\frac{\Delta k}{k}$ of
order one, and a reasonable amplitude on the order of percent are easily
obtainable within our model using values of $n\sim10^{4}$.

\section{Discussion}

\bigskip

As we have argued, chain inflation will lead to several new effects on the
spectrum of primordial perturbations. In particular, our calculations show
that the spectral index is changed from the naive one due to the presence of
interactions between radiation and the vacuum energy (the contribution
proportional to $Q_{V}\propto\varepsilon$ in formulas (\ref{spectindexeps}%
,\ref{spectindexQV})). The detailed predictions depend in a sensitive way on
the distribution of the decay rates between the different minima. If it would
be possible to measure these decay rates in a precise way, they would provide
us with a wealth of information on features of the (effective) potential of
the underlying fundamental theory. One needs to keep in mind, though, that it
is necessary to distinguish these effects from other similar effects that
could arise from non-standard potentials in other models of inflation.

Another, possibly more characteristic prediction, is the existence of
signatures similar to those that could be generated through transplanckian
physics. That is, a modulation of the spectrum due to the presence of a
fundamental scale. In case of transplanckian physics, it is the Planck scale
(or string scale) that determines the effect, while in the case of chain
inflation it is instead the size of the nucleating bubbles. It is interesting
to note that the model quite naturally, without much fine tuning, gives rise
to effects of a reasonable magnitude that possibly could be detected. As in
the case of transplanckian physics we have only been able to make a very rough
estimate of the size of the effect.

In order to make better predictions of observational signatures, a precise
model with an explicit potential and field content needs to be specified. In
\cite{Chialva:2008zw}, based on work in \cite{DanJohLar} and \cite{ChDaJoLaVo}%
, we proposed that flux compactified type IIB string theory provides such
models in a natural way. In that work we focused on the stabilization of the
complex structure moduli using fluxes.\footnote{In our simplified model the
K\"{a}hler moduli, i.e. the moduli determining the sizes of the extra
dimensions, were assumed to be fixed by other physics.} With the help of
monodromy transformations, generated by going around singular points in the
moduli space of Calabi-Yau compactifications, we were able to show the
existence of long sequences of minima of the necessary form. A quadratic
behaviour, with $c=2,$ typically arises when the axiodilaton is stabilized
independently of the complex structure moduli, while the linear behaviour with
$c=1$ arises when the axiodilation is stabilized together with the complex
structure moduli. The detailed form of the potentials depends heavily on the
choice of Calabi-Yau manifolds and fluxes, but the overall features seem to be
rather generic. In particular, the barriers in between the minima are expected
to be such that an effective slow roll behaviour arises.

It would be interesting to further explore the possibility of generating
potentials for chain inflation through string theory. Given the difficulty in
finding appropriate potentials for standard inflation, we believe this to be a
worthwhile enterprise.

\bigskip

\appendix

\section{The slow-roll parameter and its first derivative in a power-law chain
inflation model}


During inflation we naturally expect:
\begin{equation}
H^{2}\sim{8\pi G\ov3}\rho^{V} \label{approxHslow}%
\end{equation}
and\footnote{Here and in the following $\rho^{V}$ represents the energy
density of the vacuum in the interior of the bubbles. We in general expect
this formula for the following reason. First of all during inflation the total
energy density is dominated by the vacuum component. The latter is then given
by the contributions respectively of the interior of the bubbles and the
walls. But the energy density of uncollided walls is proportional to the
energy difference between two consecutive vacua, while the one of the interior
of bubbles is proportional to the energy level, which is greater than the
difference.}
\begin{equation}
\rho^{V}\sim\sum_{m}\epsilon_{m}p_{m}.
\end{equation}
Here $p_{m}(t)$ is the fraction of volume occupied by the vacuum $m$ at time
$t$ and its time evolution is given by (see \cite{Chialva:2008zw})
\begin{equation}
\dot{p}_{m}=-\widetilde{\Gamma}_{m}p_{m}+\widetilde{\Gamma}_{m+1}p_{m+1}.
\label{systemprob}%
\end{equation}
From this, we find
\begin{equation}
\dot{\rho}^{V}=-\sum_{m}\Delta\epsilon_{m}\widetilde{\Gamma}_{m}p_{m},
\end{equation}
and from (\ref{powerlaw}) we see
\begin{equation}
\Delta\epsilon_{m}\sim c{\epsilon_{m}\ov m.} \label{approxDeltaeps}%
\end{equation}
Then, using this and (\ref{approxHslow}), we find for the slow-roll parameter
\begin{equation}
\varepsilon=-{\dot{H}\ov H^{2}}={c\ov2H}\,{\sum_{m}\epsilon_{m}\widetilde
{\Gamma}_{m}p_{m}\,m^{-1}\ov\sum_{n}\epsilon_{n}p_{n}.}
\label{slowrolldiffratgenform}%
\end{equation}
If we now define an average $\langle{\widetilde{\Gamma}\ov n}\rangle$ as
\begin{equation}
\langle{\widetilde{\Gamma}\ov n}\rangle={\sum_{m}\epsilon_{m}\widetilde
{\Gamma}_{m}p_{m}\,m^{-1}\ov\sum_{n}\epsilon_{n}p_{n}},
\end{equation}
the slow-roll parameter is given by
\begin{equation}
\varepsilon\sim\frac{c}{2H}\langle{\widetilde{\Gamma}\ov n}\rangle.
\label{slowrollparneqrate}%
\end{equation}
The time derivative of the slow-roll parameter is as follows. From
(\ref{slowrolldiffratgenform}) and using (\ref{approxDeltaeps})
\begin{equation}
\dot{\varepsilon}={\varepsilon\ov2}\langle{\widetilde{\Gamma}\ov n}%
\rangle+{c\ov2H}\left(  {\sum_{m}\epsilon_{m}\widetilde{\Gamma}_{m}\dot{p}%
_{m}\,m^{-1}\ov\sum_{n}\epsilon_{n}p_{n}}+c{(\sum_{m}\epsilon_{m}%
\widetilde{\Gamma}_{m}p_{m}\,m^{-1})^{2}\ov(\sum_{n}\epsilon_{n}p_{n})^{2}%
}\right)  \label{slowrollderiv}%
\end{equation}
Let us focus on the two terms in each of the brackets. The second one is
simply
\begin{equation}
c{(\sum_{m}\epsilon_{m}\widetilde{\Gamma}_{m}p_{m}\,m^{-1})^{2}\ov(\sum
_{n}\epsilon_{n}p_{n})^{2}}=c\langle{\widetilde{\Gamma}\ov n}\rangle^{2}\,\,.
\end{equation}
It is the easy to see that using (\ref{systemprob}), the numerator in the
first term reads
\begin{align}
\sum_{m}\epsilon_{m}\widetilde{\Gamma}_{m}\dot{p}_{m}\,m^{-1}  &  =\sum
_{m}\widetilde{\Gamma}_{m+1}p_{m+1}\left(  {\epsilon_{m}\ov m}\widetilde
{\Gamma}_{m}-{\epsilon_{m+1}\ov m+1}\widetilde{\Gamma}_{m+1}\right)
\nonumber\\
&  =-\sum_{m}\widetilde{\Gamma}_{m+1}p_{m+1}\Big(\Delta\left(  {\epsilon
_{m+1}\ov m+1}\right)  \widetilde{\Gamma}_{m+1}+{\epsilon_{m}\ov m}%
\Delta(\widetilde{\Gamma}_{m+1})\Big)
\end{align}
where we have defined, for any quantity $f_{m}$
\begin{equation}
\Delta\left(  f_{m}\right)  \equiv f_{m}-f_{m-1}\,\,.
\end{equation}
We find that:
\begin{equation}
\Delta\left(  {\epsilon_{m+1}\ov m+1}\right)  =(c-1){\epsilon_{m+1}%
\ov(m+1)^{2}}%
\end{equation}
If we now define
\begin{equation}
\sigma_{\langle{\widetilde{\Gamma}\ov n}\rangle}^{2}=\left\langle \left(
\text{{\small ${\widetilde{\Gamma}\ov n}$}}\right)  ^{2}\right\rangle
-\left\langle \text{{\small ${\widetilde{\Gamma}\ov n}$}}\right\rangle
^{2}\,\,.
\end{equation}
Then, from (\ref{slowrollderiv}) and (\ref{slowrollparneqrate}), we have
\begin{equation}
\dot{\varepsilon}\sim\left(  1+{2\ov c}\right)  H\varepsilon^{2}
-(c-1)\varepsilon\,\sigma_{\widetilde{\Gamma}\ov n}^{2}\langle{\widetilde
{\Gamma}\ov n}\rangle^{-1} -\varepsilon\langle{\widetilde{\Gamma}\ov n}%
\Delta\widetilde{\Gamma}\rangle\langle{\widetilde{\Gamma}\ov n}\rangle^{-1}
+(c-1)\varepsilon\langle{\widetilde{\Gamma}\ov n}{\Delta\widetilde{\Gamma
}\ov n-1}\rangle\langle{\widetilde{\Gamma}\ov n}\rangle^{-1},
\label{derivslowrollfinalapp}%
\end{equation}
where all the averaging has been made using the distribution $\rho
_{m}=\epsilon_{m}p_{m}$.

For ease of notation, we define
\begin{equation}
D\widetilde{\Gamma}\equiv(c-1)\varepsilon\,\sigma_{\widetilde{\Gamma}%
\ov n}^{2}\langle{\widetilde{\Gamma}\ov n}\rangle^{-1}+\varepsilon
\langle{\widetilde{\Gamma}\ov n}\Delta\widetilde{\Gamma}\rangle\langle
{\widetilde{\Gamma}\ov n}\rangle^{-1}-(c-1)\varepsilon\langle{\widetilde
{\Gamma}\ov n}{\Delta\widetilde{\Gamma}\ov n-1}\rangle\langle{\widetilde
{\Gamma}\ov n}\rangle^{-1}.
\end{equation}

\bigskip

\section*{Acknowledgments}

The work was supported by the Swedish Research Council (VR) and by the EU
Marie Curie Training Site contract: MRTN-CT-2004-512194.


\bigskip

\bigskip

\begin{thebibliography}{99}                                                                                               %


\bibitem {Chialva:2008zw}D.~Chialva and U.~H.~Danielsson, ``Chain inflation
revisited,'' arXiv:0804.2846 [hep-th].

\bibitem {Freese:2004vs}K.~Freese and D.~Spolyar, ``Chain inflation: 'Bubble
bubble toil and trouble','' JCAP \textbf{0507} (2005) 007 [arXiv:hep-ph/0412145].

\bibitem {Freese:2006fk}K.~Freese, J.~T.~Liu and D.~Spolyar, ``Chain inflation
via rapid tunneling in the landscape,'' arXiv:hep-th/0612056.

\bibitem {Huang:2007ek}Q.~G.~Huang, ``Simplified Chain Inflation,'' JCAP
\textbf{0705} (2007) 009 [arXiv:0704.2835 [hep-th]].

\bibitem {Huang:2008jr}Q.~G.~Huang and S.~H.~Tye, ``The Cosmological Constant
Problem and Inflation in the String Landscape,'' arXiv:0803.0663 [hep-th].

\bibitem {kinflGarrMuk}J.~Garriga and V.~F.~Mukhanov, ``Perturbations in
k-inflation,'' Phys.\ Lett.\ B \textbf{458} (1999) 219 [arXiv:hep-th/9904176].

\bibitem {Mukhanov:1990me}V.~F.~Mukhanov, H.~A.~Feldman and
R.~H.~Brandenberger, ``Theory of cosmological perturbations.''
Phys.\ Rept.\ \textbf{215} (1992) 203.

\bibitem {Martin:2000xs}J.~Martin and R.~H.~Brandenberger, ``The
trans-Planckian problem of inflationary cosmology,'' Phys.\ Rev.\ D
\textbf{63} (2001) 123501 [arXiv:hep-th/0005209].

\bibitem {Niemeyer:2000eh}J.~C.~Niemeyer, ``Inflation with a high frequency
cutoff,'' Phys.\ Rev.\ D \textbf{63} (2001) 123502 [arXiv:astro-ph/0005533].

\bibitem {Danielsson:2002kx}U.~H.~Danielsson, ``A note on inflation and
transplanckian physics,'' Phys.\ Rev.\ D \textbf{66} (2002) 023511 [arXiv:hep-th/0203198].

\bibitem {Danielsson:2002qh}U.~H.~Danielsson, ``Inflation, holography and the
choice of vacuum in de Sitter space,'' JHEP \textbf{0207} (2002) 040 [arXiv:hep-th/0205227].

\bibitem {Easther:2002xe}R.~Easther, B.~R.~Greene, W.~H.~Kinney and G.~Shiu,
``A generic estimate of trans-Planckian modifications to the primordial
power spectrum in inflation,''
Phys.\ Rev.\ D \textbf{66} (2002) 023518 [arXiv:hep-th/0204129].

\bibitem {Bergstrom:2002yd}L.~Bergstrom and U.~H.~Danielsson,
``Can MAP and Planck map Planck physics?,''
JHEP \textbf{0212} (2002) 038 [arXiv:hep-th/0211006].

\bibitem {Coleman:1977py}S.~R.~Coleman, ``The Fate Of The False Vacuum. 1.
Semiclassical Theory,'' Phys.\ Rev.\ D \textbf{15} (1977) 2929
[Erratum-ibid.\ D \textbf{16} (1977) 1248].

\bibitem {Coleman:1980aw}S.~R.~Coleman and F.~De Luccia, ``Gravitational
Effects On And Of Vacuum Decay,'' Phys.\ Rev.\ D \textbf{21} (1980) 3305.

\bibitem {DanJohLar}U.~H.~Danielsson, N.~Johansson and M.~Larfors, ``The world
next door: Results in landscape topography,'' JHEP \textbf{0703} (2007) 080 [arXiv:hep-th/0612222].

\bibitem {ChDaJoLaVo}D.~Chialva, U.~H.~Danielsson, N.~Johansson, M.~Larfors
and M.~Vonk, ``Deforming, revolving and resolving - New paths in the string
theory landscape,'' JHEP \textbf{0802} (2008) 016 [arXiv:0710.0620 [hep-th]].

\bibitem {Hamann:2008yx}J.~Hamann, S.~Hannestad, M.~S.~Sloth and
Y.~Y.~Y.~Wong, ``Observing trans-Planckian ripples in the primordial power
spectrum with future large scale structure probes,'' arXiv:0807.4528 [astro-ph].
\end{thebibliography}
\end{document}